\documentclass[nohyper,12pt,letterpaper]{JHEP3}

\usepackage{amsmath,epsfig}
\usepackage[latin1]{inputenc}
\usepackage{amssymb,amsfonts}
\usepackage{cite}

\title{BPS Black Hole Degeneracies \\
and Minimal Automorphic Representations}

\preprint{\hepth{0506228}\\LPTHE-05-15\\LPTENS-05-20}

\author{Boris~Pioline\\\\
LPTHE, Universit\'es Paris 6 et 7, \\ 4 place
Jussieu,75252 Paris cedex 05, France\\ \\
and\\ \\
LPTENS, D\'epartement de Physique de l'ENS, \\
24 rue Lhomond, 75231 Paris cedex 05, France\\ \\
E-mail: {\tt pioline@lpthe.jussieu.fr}}

\abstract{We discuss the degeneracies of 4D and 5D BPS black holes
in toroidal compactifications of M-theory or type II string
theory, using U-duality as a tool. We generalize the 4D/5D lift to include
all charges in $\CN=8$ supergravity, and compute the exact indexed
degeneracies of certain 4D $1/8$-BPS black holes. 
Using the attractor formalism, we obtain the 
leading micro-canonical entropy for arbitrary Legendre invariant prepotentials
and non-vanishing D6-brane charge.
In particular, we find that the $\CN=8$ prepotential 
is given to leading order 
by the cubic invariant of $E_6$. This suggests that the
minimal unitary representation of $E_8$, based on the same cubic prepotential,
underlies the microscopic degeneracies of $\CN=8$ black holes. We propose
that the exact degeneracies are given by the Wigner function 
of the $E_8(\IZ)$ invariant vector
in this automorphic representation. A  similar conjecture relates 
the degeneracies of $\CN=4$ black holes
to the minimal unipotent representation of $SO(8,24,\IZ)$.}

\renewcommand{\subsubsection}{\@startsection{subsubsection}{3}{0mm}{-\baselineskip}{0.5\baselineskip}{\normalfont\normalsize\it}}

\renewcommand{\Im}{{\rm Im}}

\newcommand{\pa}{\partial}

\newcommand{\IR}{\mathbb{R}}
\newcommand{\IZ}{\mathbb{Z}}

\newcommand{\Tr}{\mbox{Tr}}
\newcommand{\Pf}{\mbox{Pf}}

\newcommand{\CN}{{\cal N}}
\newcommand{\CX}{{\cal X}}
\newcommand{\CH}{{\cal H}}

\def\bea{\begin{eqnarray}}
\def\eea{\end{eqnarray}}
\def\be{\begin{equation}}
\def\ee{\end{equation}}
\def\ba{\begin{align}}
\def\ea{\end{align}}
\def\bse{\begin{subequations}}
\def\ese{\end{subequations}}

\begin{document}

\section{Introduction}
One of the distinct successes of string theory is to provide a statistical
interpretation of the Bekenstein-Hawking entropy of a class of extremal or
near-extremal dyonic black holes, in terms of manifestly unitary micro-states 
\cite{Strominger:1996sh,Callan:1996dv,Maldacena:1996gb,Johnson:1996ga}. 
While this agreement was originally obtained in 
the limit of large electric and magnetic charges, corresponding to large
horizon area in Planck units, subleading corrections to the entropy 
have received renewed attention recently 
\cite{Ooguri:2004zv,Dabholkar:2004yr,Cardoso:2004xf,Sen:2004dp,Sen:2005ch,
Sen:2005pu,Sen:2005kj,Verlinde:2004ck,Dabholkar:2005by,
Ooguri:2005vr,Dijkgraaf:2005bp,Kraus:2005vz,Sen:2005wa} 
(see \cite{LopesCardoso:1998wt,LopesCardoso:1999cv,
LopesCardoso:1999ur,LopesCardoso:1999xn,Maldacena:1997de} for early
studies). On the macroscopic side, the latter arise from 
higher-derivative interactions in the effective 
action \cite{Wald:1993nt,Iyer:1994ys}, while on the microscopic side,
they depend on the fine details of the underlying quantum mechanics,
including a choice of statistical ensemble. 
Based on a re-interpretation of the attractor 
mechanism \cite{Ferrara:1995ih,Ferrara:1996dd,Ferrara:1996um},
suitably generalized to include a class of `F-type'' 
interactions\cite{LopesCardoso:1998wt,LopesCardoso:1999cv,
LopesCardoso:1999ur,LopesCardoso:1999xn}, these subleading corrections
to the macroscopic entropy have been conjectured to reflect 
finite size corrections to the microscopic entropy
in a specific ``mixed'' statistical ensemble \cite{Ooguri:2004zv}.
Furthermore, it has become apparent that the Bekenstein-Hawking-Wald 
entropy may be protected from ``non-F-type'' contributions,
at least of a particular class of 
BPS black holes \cite{Kraus:2005vz,Sen:2005wa}.
Independently of these developments,  a precise connection 
between 4D black holes, 5D black holes and 5D black strings has 
begun to emerge \cite{Gaiotto:2005xt,Gaiotto:2005gf,
Elvang:2005sa}, providing a new handle on the counting
of black hole micro-states \cite{Dijkgraaf:1996it,Shih:2005uc}.

It is therefore of interest to reconsider the entropy 
of BPS black holes in maximally supersymmetric theories, where 
U-duality \cite{Hull:1994ys} 
is expected to provide a powerful constraint on the 
higher-derivative terms in the effective action, as well as on the microscopic
degeneracies (see \cite{Obers:1998fb} for a review of U-duality). 
The indexed degeneracies of 5D 1/4-BPS black holes in type II 
string theory compactified on $T^5$ (or M-theory
on $T^6$) were computed in \cite{Maldacena:1999bp}, relying on the
invariance under the U-duality group $E_6(\IZ)$. On the 
other hand, the leading Bekenstein-Hawking entropy of 1/8-BPS 4D black
holes in type II compactified on $T^6$  
(or M-theory on $T^7$) is known to be controlled by the quartic invariant
of $E_7$ \cite{Kallosh:1996uy,Ferrara:1996um,Cvetic:1996zq}. 
The aim of this
work is to determine the subleading corrections to this formula, and
formulate a conjecture which relates the exact (indexed)
degeneracies of 4D 1/8-BPS black holes
to automorphic representations of the U-duality group.

A brief outline of this work is as follows. In Section 2, we review
some relevant facts about M-theory compactified on $T^7$ and $T^6$, with
special emphasis on U-duality. In particular, we introduce an important
relation \eqref{i4i3} between the quartic invariant of $E_7$ and
the cubic invariant of $E_6$, which plays a central r\^ole in the sequel.

In Section 3, we combine the 4D/5D lift of \cite{Gaiotto:2005xt}
and the 5D counting of \cite{Maldacena:1999bp} to obtain the exact 
helicity supertrace $\Omega_8$ of the micro-states of 
four-dimensional 1/8-BPS black holes. Based on the relation
\eqref{i4i3} (or rather its equivalent form \eqref{i4i3t}), 
we obtain in \eqref{gen4d5d} a generalization of the 4D/5D lift to 
all charges in $\CN=8$ supergravity.

In Section 4, we compute the micro-canonical degeneracies predicted by
the conjecture in \cite{Ooguri:2004zv}, for general Legendre invariant
tree-level prepotentials $F_0=I_3(X)/X^0$ and arbitrary electric
and magnetic charges (including the D6-brane charge), in the semi-classical
approximation. The assumption of Legendre invariance greatly simplifies
the computation, and is in fact a property of the prepotentials describing
homogeneous vector-multiplet moduli spaces \cite{Gunaydin:1983bi}.
In particular, we find that the $E_7$-invariant 
Bekenstein-Hawking entropy is correctly reproduced as a function
of all charges, provided $I_3(X)$ is chosen to be the cubic invariant
of $E_6$. This relies crucially on the relation \eqref{i4i3}, and in
fact provides a derivation (or rationale) of Eqs.\ \eqref{i4i3},\eqref{i4i3t}.
We thus conclude that the topological amplitude in $\CN=8$ string
theory is, to leading order, $\Psi=\exp(I_3(X)/X_0)$, where 
$I_3(X)$ is the cubic invariant of $E_6$.

At this point, we observe that this $E_6$-invariant prepotential 
also underlies the minimal unipotent representation of $E_8(\IR)$ 
constructed in \cite{Kazhdan:2001nx,Gunaydin:2001bt}. This is
a unitary representation of $E_8$ acting on a Hilbert 
space $\CH$ of functions of 29 variables, 28 of which can be
understood as the 28 electric charges of $\CN=8$ supergravity.
This suggests that the degeneracies of 4D 1/8-BPS black holes 
may have a hidden $E_8(\IZ)$ symmetry, upon including an extra
quantum number.
The idea that $E_8$ may act as a ``spectrum generating'' symmetry
has been suggested in the past \cite{Ferrara:1997uz,Gunaydin:2000xr,
Gunaydin:2001bt}, and is quite natural given that black holes 
in 4 dimensions can be viewed as instantons in 3 Euclidean dimensions,
where the U-duality group is enlarged to $E_8(\IZ)$
(for an analogous reason, the entropy of 5D black rings
exhibits a hidden $E_7$ symmetry \cite{Bena:2004tk}).
The fact that certain partition functions have a higher degree of 
symmetry than expected is also familiar in toroidal
string compactifications (where the product of the T-duality
group $SO(d,d,\IZ)$ and genus $g$ modular group
$Sp(2g,\IZ)$ are embedded in a larger symplectic group $Sp(2gd,\IZ)$,
which is a symmetry of the partition function of the bosonic 
zero-modes \cite{Obers:1999um}) and in membrane theory 
(where the  product of the 1-loop modular group $Sl(3,\IZ)$ 
and the U-duality group $E_d(\IZ)$ are embedded in
a larger $E_{d+2}(\IZ)$, conjectured to be a symmetry of
the BPS membrane partition function \cite{Pioline:2001jn,Pioline:2004xq}). 

In Section 5, we try and flesh out this idea. 
After a brief review of the general construction of minimal
representations, we identify the 29 variables
in the minimal representation of $E_8$ as the
28 electric charges together with the NUT charge which arises
in the reduction to three dimensions along the time direction.
By analogy with the metaplectic representation of $Sl(2)$,
which we recall in Subsection 5.3,
we propose that the black hole 
degeneracies are given by the Wigner function of a $E_8(\IZ)$ invariant
distribution in $\CH$. As explained in \cite{Kazhdan:2001nx,Pioline:2003bk}, 
this distribution is the measure 
for the non-gaussian theta series of $E_8$, and is the product over all
primes $p$ of the spherical vector of the representation over
the $p$-adic numbers $\mathbb{Q}_p$. We sketch a similar conjecture
for 1/4-BPS black holes in heterotic string compactified on $T^6$
(or type II string theory compactified on $K3\times T^2$), 
which we argue is related to the 
minimal representation of the 3D U-duality group $D_{16} = SO(8,24)$. 
Finally, we suggest that the conformal quantum mechanics which
underlies the minimal representation of $E_8$ 
\cite{Gunaydin:2001bt,Pioline:2002qz} may be the $\CN=8$
realization of the quantum cosmology / attractor flow scenario
considered in \cite{Ooguri:2005vr,Dijkgraaf:2005bp}.

Admittedly, the conjectures in Section 5 
are rather speculative, and it would be 
very desirable to understand the relation with earlier proposals 
such as \cite{Dijkgraaf:1996it,Shih:2005uc} in the $\CN=4$ case,
or \cite{Dijkgraaf:1996cv,Dijkgraaf:1996hk,Maldacena:1999bp} in the 
$\CN=8$ case. If correct, their generalization to $\CN=2$ zupersymmetry may
turn out to have very interesting mathematical consequences.

For completeness, in Appendices A and B we discuss
some possible applications of the minimal automorphic representations 
of $E_7(\IZ)$ and $E_6(\IZ)$ to 5D and 6D black holes, respectively.

\section{Black hole entropy and U-duality} 
Let us start by recalling a few relevant 
facts about M-theory compactified on $T^7$.
The massless spectrum in 4 dimensions consists of the graviton, 8 gravitini,
28 abelian gauge fields, 56 fermions and 70 scalars. The 70=28+35+7
scalars come
from the reduction of the 11 dimensional metric $g_{IJ}$, 3-form $C_{IJK}$
and 6-form $E_{IJKLMN}$ (the dual of the 3-form in 11 dimensions) on
$T^7$, respectively,
and parameterize the  symmetric space $E_7/SU(8)$ \cite{Cremmer:1978ds}.
The 28 gauge fields together with their magnetic duals transform into a 56
representation of $E_7$. They arise by reduction of the above 11-dimensional
fields, together with $K_{I;JKLMNPQR}$, which represents the magnetic dual
of the graviton \cite{Obers:1998fb}\footnote{$K_{I;JKLMNPQR}$
transforms as $\Lambda^1 \otimes \Lambda^8$ under $Sl(11)$,
and is best thought of as 
the multiplet of Kaluza-Klein gauge fields $g_{I\mu}$ after reduction to
3 dimensions.}:
\be
\label{gf4}
7\ g_{\mu I},\ 
21\ C_{\mu IJ},\ 
21\ E_{\mu IJKLM},\
7\ K_{I;\mu JKLMNPQ}
\ee
The corresponding charges can be fit into two $8\times 8$ antisymmetric
matrices,
\be
\label{QPmat}
Q=\begin{pmatrix} [M2]^{IJ} & [KKM]^I \\ -[KKM]^I & 0 \end{pmatrix}\ ,\quad
P=\begin{pmatrix} [M5]_{IJ} & [KK]_I \\ -[KK]_I & 0 \end{pmatrix}
\ee
where $[KK]_I$ corresponds to a momentum excitation along the compact
direction $I$, $[M2]^{IJ}=-[M2]^{JI}$ to 
a M2-brane wrapped on the directions $IJ$, $[M5]_{IJ}=-[M5]_{JI}$ to a M5-brane
wrapped on all compact directions {\it but} $IJ$, and $[KKM]^I$ to
a Kaluza-Klein monopole wrapped in all compact directions but $I$. This
splitting into ``electric'' charges $Q$ and ``magnetic'' charges $P$ is not
the usual ``large volume'' polarization, but it is the one that makes
the $Sl(8)$ subgroup of the $E_7$ symmetry manifest.

The Bekenstein-Hawking entropy of 1/8-BPS black holes
is given by  \cite{Kallosh:1996uy,Ferrara:1996um,Cvetic:1996zq}
\be
\label{kol}
S_{BH,4D}= \pi \sqrt{I_4(P,Q)}
\ee
where $I_4(P,Q)$ is the singlet in the symmetric tensor
product of four 56 of $E_7$, also known as the ``diamond''
invariant:
\be
\label{i4}
I_4(P,Q)= -\Tr(QPQP) 
+ \frac14 \left( \Tr QP \right)^2 
-4 \left[ \Pf(P) + \Pf(Q)  \right]
\ee
(The Pfaffian is, as usual, the square root of the determinant 
of an antisymmetric matrix; the choice of branch is purely conventional).

Viewing the direction 1 as the dynamically generated dimension
of type IIA string theory compactified on $T^6$,
and taking the weak string coupling, the moduli space decomposes in 
regime into a product
\be
\label{e7so66}
\frac{E_7}{SU(8)} = 
\frac{Sl(2)}{U(1)} \times \frac{SO(6,6)}{SO(6)\times SO(6)} \bowtie \IR^{32}
\ee
The first and second factor describe the axio-dilaton $E_{234567} + 
i V_{234567}/g_s^2 l_s^6$ and the Narain moduli of $T^6$,
respectively, and correspond to a $\CN=4$ supersymmetric truncation of 
the spectrum. The third factor corresponds
to the Ramond-Ramond gauge potentials on $T^6$, and 
transforms as a spinor representation of the T-duality group $SO(6,6)$.
The black hole charges decompose into $(2,12) \oplus (1,32)$ 
under $Sl(2)\times SO(6,6)$, corresponding to 6 Kaluza-Klein 
momenta $[kk]_i=[KK]_i$,
6 fundamental string windings $[F1]^i=[M2]^{1i}$, 32 wrapped D-branes 
$[D0]=[KK]_1,[D2]^{ij}=[M2]^{ij}, [D4]_{ij}=\epsilon_{ijklmn}
[D4]^{klmn}/6
=[M5]_{ij},
[D6]=[KKM]^1$, 6 wrapped NS5-branes $[NS]_{i}=[M5]_{1i}$ and 
6 wrapped KK5-monopoles $[kkm]^i=[KKM]^i$ (here $i,j=2,\dots,7$):
\be
Q=\begin{pmatrix} 
[D2]^{ij} & [F1]^i & [kkm]^i \\ 
-[F1]^i & 0 & [D6]  \\ 
- [kkm]^i & - [D6] &  0
\end{pmatrix}\ ,\quad
P=\begin{pmatrix} 
[D4]_{ij} & [NS]_i & [kk]_i \\ 
-[NS]_i & 0 & [D0]  \\ 
- [kk]_i & - [D0] &  0
\end{pmatrix}\ ,\quad
\ee
The $\CN=4$ truncation keeps the string winding, momenta, NS5-brane 
and KK5-monopoles, but throws away the D-branes. It is easy to check
that the entropy formula \eqref{kol} reduces to the standard $\CN=4$ 
answer \cite{Cvetic:1995bj,Ferrara:1996um},
\be
S = \pi \sqrt{ (\vec q_e)^2 (\vec q_m)^2 - (\vec q_e \cdot \vec q_m)^2}
= \pi \sqrt{q^I_\alpha\ q^J_\beta\ q^K_\gamma\ q^L_\delta\ 
\eta_{IK}\ \eta_{JL}\ \epsilon^{\alpha\beta}\ \epsilon^{\gamma\delta} }
\ee
where $\eta_{ij}$ is the signature (6,6) metric and 
$q^I_1= ([kk]_i,[F1]^i),  q^I_2= ([NS]_i, [kkm]^i)$.

It is also of interest to discuss the strong coupling limit
where the direction 1 decompactifies, leading to 
M-theory compactified on $T^6$, with a U-duality group $E_6$. The multiplet of 
4D black hole charges decomposes under $E_6$ into $1+27+\overline{27}+1$ 
charges ($i,j=2\dots 7$)
\bse
\label{lr}
\be
\label{lr1}
q_0=[KK]_1\ ;\qquad Q_A= \{ [M2]^{ij}\ ,\quad [KK]_i\ ,\quad [M5]_{i1} \}
\ee
\be
\label{lr2}
p^0=[KKM]^1\ ;\qquad P^A= \{ [M2]^{i1}\ ,\quad [KKM]^i\ ,\quad [M5]_{ij} \}
\ee
\ese
where the $27$ charges $q_A$ in the 
first line correspond to 5D black holes, while the $\overline{27}$
charges $p^A$ in the second line correspond to 5D black strings 
wrapped along direction 1 (or dipole charges), which become infinitely
massive in the strict infinite coupling limit. This splitting
agrees with the one corresponding to the large volume limit
of type IIA string on $T^6$,
\bse
\label{lv}
\be
\label{lv1}
q_0=[D0]\ ; \qquad q_A = \{ [D2]^{ij}\ ,\quad [kk]_i\ ,\quad [NS]_i \}
\ee
\be
\label{lv2}
p^0=[D6]\ ;\qquad p^A = \{ [D4]_{ij}\ ,\quad [F1]^i\ ,\quad [kkm]^i \}
\ee
\ese
Nevertheless, for reasons which will become clear below, it is useful
to use a different symbol for the 5D black hole charges $Q_A$ and the 
4D electric charges $q_A$ (and similarly, for the 5D dipole charges $P^A$
and the 4D magnetic charges $p^A$). 
The entropy of the 5D black holes is then given 
by \cite{Ferrara:1996dd,Dijkgraaf:1996cv}
\be
\label{s5d}
S_{BH,5D} = 2\pi \sqrt{I_3(Q_A) - (J_L^3)^2 }
\ee
where $I_3$ is the cubic invariant of $E_6$,
\be
I_3(Q_A) = \Pf \left( [M2]^{ij} \right) + \frac{1}{5!} \epsilon_{jklmnp} 
[KK]_i [M2]^{ij} [M5]^{klmnp}
\ee
and $J_L^3$ is the angular momentum in 5 dimensions.

A central observation for the sequel is that, in the large volume
basis \eqref{lv}, the $E_7$ quartic invariant \eqref{i4}
can be expressed in terms of the $E_6$ cubic invariant 
as follows\footnote{This relation is in fact known to arise
in Freudenthal's triple system
construction of exceptional groups, see e.g. Eq. (2.15) in
\cite{Ferrara:1997uz} and references therein.}:
\be
\label{i4i3}
I_4(p,q)= 4 p^0 I_3(q_A) - 4 q_0 I_3(p^A) + 4  
\frac{\pa I_3(q_A)}{\pa q_A}  \frac{\pa I_3(p^A)}{\pa p^A} 
- ( p^0 q_0 + p^A q_A )^2
\ee
where $q_A$ and $p^A$ ($A=1,\dots,27$) 
are the 27 and $\overline{27}$ multiplets
in \eqref{lv1}, \eqref{lv2}, and $q_0,p^0$ are the D0 and D6-brane charge. This
equation may be easily checked by explicit computation, but, as we shall 
demonstrate in Section 4.2, it is a general consequence of the invariance
of the $\CN=8$ prepotential $F_0=I_3(X)/X^0$ under Legendre transform.
It is also usefully rewritten as 
\be
\label{i4i3t}
I_4(p,q)=  \frac{1}{(p^0)^2}\left[ 4 I_3(Q_A) - (2 I_3(p^A) + p^0 p^I
  q_I)^2 \right]
\ee
where the sum over $I$ runs from 0 to 27, and, intentionally
using the same notation as in \eqref{lr}, 
\be
Q_A = p^0 q_A + \partial_A I_3(p^A)\ .
\ee
As we shall see in Section 3.2, this version
of the identity embodies the 4D/5D lift of \cite{Gaiotto:2005gf} 
generalized to all charges of $\CN=8$ supergravity.

Finally, let us discuss the $\CN=2$ truncation of this theory.
It is well known that the $\CN=8$ gravity multiplet
splits into 1 $\CN=2$ gravity multiplet, 6 $\CN=2$ gravitini multiplets, 
15 vector multiplets and 10 hypermultiplets \cite{Gunaydin:1983rk}.
We are interested in a truncation which preserves the general 
structure of type IIA compactifications on a Calabi-Yau three-fold $\CX$,
where vector multiplets arise from two-cycles in $H_{1,1}(\CX)$. 
Since $H_2(T^6)=\IZ^{9}$, we are interested in a truncation which
keeps only the 9 vector multiplets. This corresponds to the $T^6/\IZ_3$
orbifold \cite{Ferrara:1988fr}, with prepotential
\be
F_0 = \frac{\det(X)}{X^0}
\ee
where $X_{i\bar j}$ is the  $3\times 3$ complex matrix of K\"ahler
moduli. The resulting scalar manifold is the symmetric space
$SU(3,3)/S(U(3)\times U(3))$.

However, due to the flatness of $T^6$, there also exist BPS branes wrapped on
two-cycles in $H_{2,0}$ and $H_{0,2}$: it is thus natural to treat all 
2-cycles in $H_2(T^6)=\IZ^{15}$ at once, and consider the 
generalized prepotential 
\be
\label{f0n2}
F_0 = \frac{\Pf(X)}{X^0}
\ee
where $X$ is now a $6\times 6$ antisymmetric matrix of complex
moduli\footnote{
Equivalently, the 15 complex moduli may be fit into a $3\times 3$
hermitian matrix with quaternionic coefficients, whose determinant
is equal to $\Pf(X)$.}, resulting in the symmetric 
space $SO^*(12)/U(6)$ \cite{Gunaydin:1983rk}. The cubic
polynomial $\Pf(X)$ is recognized as the cubic intersection form on $H_2(T^6)$.
The corresponding electric and magnetic charges are 
all 32 D-brane charges $[D0],[D2]^{ij},[D4]_{ij},[D6]$, transforming as
a spinor of $SO(6,6)$.
Setting $[kk]=[kkm]=[F1]=[NS]=0$, the entropy formula \eqref{kol} 
truncates to $S_{BH;4D}= \pi \sqrt{\tilde I_4}$
where 
\be
\label{it4}
\tilde I_4
= 4 [D6] \Pf([D2]) - 4 [D0] \Pf([D4]) + 4 \Tr( [D2] [D4] [D2] [D4] ) - 
( [D0][D6] + [D2][D4] )^2\ ,
\ee
which is recognized as the singlet in the symmetric tensor product
of 4 spinor representations of $SO(6,6)$. It is worth mentioning that 
formula \eqref{i4i3} still holds upon replacing $I_4$ by $\tilde I_4$
and $I_3$ by $\tilde I_3(q) = \Pf(X)$.

Since the moduli space is not corrected due to $\CN=8$ supersymmetry,
the tree-level prepotential \eqref{f0n2} is in fact exact. 
Note also that the higher genus topological
amplitudes $R^2 F^{2h-2}$ vanish. However, it is conceivable that
higher-derivative $R^4 H^{4h-4}$ interactions, computed by 
the $\CN=4$ topological string \cite{Berkovits:1998ex}, 
may contribute to the topological
amplitude in the $\CN=8$ setting.

\section{Exact degeneracies of 1/8-BPS states}
In this section, we combine the 4D/5D lift of \cite{Gaiotto:2005gf} 
with the degeneracies of 
5D black holes computed in \cite{Maldacena:1999bp} 
to derive the exact (indexed) degeneracies of 
a class of 4D 1/8-BPS black holes.

\subsection{1/8-BPS states in II/$T^5$}

Let us start by reviewing the result of \cite{Maldacena:1999bp},
who computed a particular index 
\be
\label{ind5}
\Omega_{5D} = \Tr (-1)^{2 J^3_L - 2 J^3_R} (2J_3^R)^2
\ee
in the Hilbert space of BPS black holes in type IIB string theory compactified 
on $T^5=T^4\times S^1$, with fixed electric charges $Q_A\in 27$ and 
angular momentum $\ell=2J_3^L$.
By a U-duality rotation, one may choose the standard configuration of 
$Q_1$ D1-branes wrapping $S^1$, $Q_5$ branes wrapping $S^1\times T^4$
and $N$ units of momentum along the circle $S_1$.  By analysing the
generalized elliptic genus of $Hilb(T^4)$, the  authors 
of \cite{Maldacena:1999bp} conjectured the relation
\be
\label{om55}
\Omega_{5D}(N,Q_1,Q_5,\ell) = \sum_{s|(NQ_1,NQ_5,Q_1Q_5,\ell);s^2|NQ_1Q_5} 
\ s \ N(s)\  \hat c\left( 
\frac{N Q_1 Q_5}{s^2} , \frac{\ell}{s} \right)
\ee
where $N(s)$ is the number of divisors of
\be
\label{gra}
N , Q_1, Q_5, s, \frac{NQ_1}{s}, \frac{NQ_5}{s}, \frac{Q_1 Q_5}{s}, 
\frac{NQ_1 Q_5}{s^2}, 
\ee
$\hat c(n,l)$ are the Fourier coefficients of the weak Jacobi form
\be
- \frac{\theta_1^2(z,\tau)}{\eta^6} := \sum_{n=0}^{\infty}
\sum_{l\in\IZ}
 \hat c(n,l) q^n y^l 
\ee
This formula was rigorously established for $N,Q_1,Q_5$ coprime, and
is manifestly invariant under the subgroup of $E_6(\IZ)$ which permutes
$N,Q_1,Q_5$.

Since $Z$ is a weak Jacobi form of weight -2 and index 1, the Fourier
coefficients are function of a single variable,
\be
\hat c(n,l)=\hat c(4n-l^2)
\ee
with $\hat c(-1)=1, \hat c(0)=-2, \hat c(1)=8, \hat c(4) = -12, \dots$.
In fact, the generating function of these coefficients is a simple
modular form
\be
\label{hatcq}
\Phi(\tau)= \sum_{n=-1}^{\infty} 
\hat c(n) q^n  = \frac{\theta_4(2\tau)}{\eta^6(4\tau)} = 
\frac{2^4}{\theta_2^4(2\tau) \theta_3(2\tau)}
\ee
whose Fourier coefficients can be approximated to great accuracy by 
the Rademacher formula \cite{Dijkgraaf:2000fq,Dabholkar:2005by}. 
Restricting for simplicity to the case 
where $N,Q_1,Q_5,\ell$ are coprime, we find
\be
\label{om5}
\Omega_{5D} \sim \hat I_{7/2} \left( \pi \sqrt{ 4 N Q_1 Q_5 -\ell^2 }
\right)\ ,
\ee
up to computable exponentially suppressed corrections.
Using the usual asymptotic expansion of the modified 
Bessel function $\hat I_\nu(z)$ \cite{Dabholkar:2005by}, we find 
\be
\ln \Omega_{5D} = 2\pi   \sqrt{ N Q_1 Q_5 -J_L^2 } 
- 4 \log ( N Q_1 Q_5 - J_L^2 )  + \dots
\ee
In particular, this formula predicts an infinite number of subleading 
corrections to the tree-level Bekenstein-Hawking entropy \eqref{s5d}.
It would be interesting to relate these corrections to higher-derivative
couplings in the effective action such as $R^4$.

\subsection{From 4D to 5D black holes}
We now apply the relation between 4D and 5D black holes 
established recently in \cite{Gaiotto:2005xt}: a 4D black hole in IIA/CY 
with charges $[D6], [D2]_{ij}, [D0]$ 
and but no D4-charge is equivalent to a 5D black hole
with M2-brane charge $[M2]^{ij}=[D6] [D2]^{ij}$ and 
angular momentum $2J_L^3 = [D6]^2[D0]$,
at the tip of a Taub-NUT gravitational instanton with charge $p^0=[D6]$. 
Since the geometry at the tip is locally $\IR^4/\IZ_{p^0}$, 
the Bekenstein-Hawking entropy of the 4D black hole \eqref{kol}
should be given by $1/p^0$ times the Bekenstein-Hawking entropy \eqref{s5d}
of the 5D black holes. Indeed, for the above choice of charges,
\be
\label{s4d}
S_{BH;4D} = 2\pi \sqrt{ [D6] \Pf([D2]) - \frac14 ([D0][D6])^2 } 
= \frac{2\pi}{p^0} \sqrt{ \Pf([M2]) - J_L^2 } 
= \frac{1}{p^0} S_{BH;5D} 
\ee
As a matter of fact, this observation can be generalized to the $\CN=8$
setting, by using the identity \eqref{i4i3t} to rewrite the 
4D black hole entropy as
\be
S_{BH;4D} = \frac{\pi}{|p^0|}
\sqrt{4 I_3(Q_A) - (2 I_3(p^A) + p^0 p^I  q_I)^2}
\ee
The Bekenstein-Hawking of the
4-dimensional black hole 
is thus equal to  $1/p^0$ times the Bekenstein-Hawking entropy 
of a 5-dimensional black hole \eqref{s5d}
provided the charges are identified as
\bse
\label{gen4d5d}
\be
Q_A = p^0 q_A + \partial_A I_3(p)\ ,\quad 
\ee
\be
2J_L =  (p^0)^2 q_0 + p^0  p^A q_A + 2 I_3(p) 
\ee
\ese
In more detail, 
\bse
\be
[M2]^{ij}=[D6] [D2]^{ij} + \frac18 \epsilon^{ijklmn} [D4]_{kl} [D4]_{mn}
+ [F1]^i[kkm]^j-[F1]^j[kkm]^i 
\ee
\be
[KK]_{i}=[D6][kk]_i + [D4]_{ij} [kkm]^j
\ee
\be
[M5]_{i}=[D6][NS]_i + [D4]_{ij} [F1]^j 
\ee
\be
2J_L^3 = [D6] ( [D6] [D0] + \frac12 [D4]_{ij} [D2]^{ij} 
+ [NS]_i[F1]^i+[kk]_i [kkm]^i )
\ee
\ese
It would be interesting to support this algebraic observation 
by a construction of the actual supergravity solutions.

For $[D6]>1$, the orbifold singularity at the tip of the cigar implies that
the 4D black hole will have additional twisted micro-states compared to the
5D one, which will affect subleading corrections to the entropy. For $[D6]=1$
however, one can assume that these effects are absent and directly obtain
the exact degeneracies of 4D black holes from the corresponding 5D 
black hole \cite{Gaiotto:2005gf,Shih:2005uc}.

Following \cite{Shih:2005uc}, consider now a 4D black hole
in type II compactified on $T^6=T^4\times T^2$ 
with $q_0$ D0-branes , $q_1=[D2]^{12}$ D2-branes wrapped on $T^2$,
$q^{ab}=-q^{ba}$ D2-branes wrapped on $T^4$ and one unit of D6-brane charge.
This lifts to a 5D black hole 
in M-theory on $T^4\times T^2$ with spin $J_L=q_0/2$ and M2 charge
$[M2]^{ij}=(q_{1},q_{ab})$. Identifying one of the circles on $T^2$
as the M-theory circle, this is equivalent to 
5D black hole in IIA string theory compactified $T^4\times S^1$ 
with $q_{1}$ F1-strings, $q^{ab}$ 
D2-branes wrapping $T^4$ and the same spin $J_L=q_0/2$. 
By a sequence of T-dualities,
this is mapped to the standard D1-D5-kk system in 
type IIB/ $T^4\times S^1$, with central charge $c$, angular momentum 
$J_L^3$ 
and left-moving momentum $L_0$ along $S^1$ given by 
\be
c= 6\Pf (q_{ab})\ , \quad
J_L^3 = \frac12 q_0\ ,\quad
L_0 =q_{1}
\ee
The five-dimensional index \eqref{ind5} is further identified to 
\be
\label{ind5b}
\Tr' (-1)^{2J_3} (2J_3)^2 
\ee
where $J_3$ is the Cartan component of the 4-dimensional spin and 
$\Tr'$ denotes the trace with the center of mass multiplet factored 
out \cite{Gaiotto:2005gf}. Reinstating the center of mass coordinates,
we find that \eqref{ind5b} computes the eighth helicity supertrace $\Omega_8$
in 4 dimensions, which is the first non-vanishing supertrace for $1/8$-BPS
multiplets (\cite{Kiritsis:1997gu}, Appendix G).
According to \eqref{om55}, the exact indexed degeneracy is therefore 
\be
\Omega_{8} = \sum_s  s\ N(s)\ \hat c\left( \frac{\Pf (Q)}{s^2} ,  
\frac{q_0}{s}\right)
\ee
where $\Pf(Q)= q_{1} \Pf(q)$ and $\hat c(n)$ are the Fourier coefficients
of the modular form in \eqref{hatcq}. Assuming that all charges are coprime,
using \eqref{om5} we find that the microscopic degeneracies grow as
\be
\label{smic}
\Omega_{8} \sim \hat I_{7/2} \left( \pi \sqrt{ I_4} \right)
\ ,\quad I_4 = 4 \Pf(Q) - q_0^2 
\ee
Using the generalized 4D/5D lift in  \eqref{gen4d5d}, it is natural 
to conjecture that, more
generally, the eighth-helicity supertrace should be given by
\be
\Omega_{8} = \sum_{s : \nabla_X F \in \IZ} 
\ s \ N(s)\  \hat c\left( 
\frac{I_3(Q^A)}{s^2} , \frac{J_L}{s} \right)
\ee
where $Q^A$ and $J_L$ are given in  \eqref{gen4d5d},
and $N(s)$ is the number of common divisors of
$X^I$ and $\nabla_I F_0$, where 
\be
F_0 = \frac{I_3(X^A)}{X^0}\ ,
\quad X = (s ; [D2]^{ij}, [NS]_i, [kk]_i) 
\ee
The sum over $s$ should of course be restricted to values such
that all $X$ and $\nabla_X F_0$ be integers. In addition, $s$ should also
divide $J_L/2$. Thanks to \eqref{i4i3t}, this proposal clearly reproduces
the correct leading entropy. Unfortunately due to the existence
of twisted sectors when $p^0\neq 1$, it is unclear that the 
subleading contributions are correctly predicted. In Section 5.3, we will
formulate a conjecture which potentially predicts the exact degeneracies
of all 1/8-BPS states.

\section{Comparison to the topological string amplitude}
In general, we expect that the subleading
contributions in the microscopic entropy should be related to 
corrections to the macroscopic Bekenstein-Hawking entropy, 
due to higher-derivative interactions in the effective action.
A immediate problem with this idea is that subleading corrections
to the entropy are non-universal, and depend on a choice of statistical
ensemble. In models with $\CN=2$ supersymmetry, it has been suggested that
the appropriate ensemble to match the macroscopic answer should be a 
``mixed'' ensemble where magnetic charges $p^I$ are treated micro-canonically,
whereas electric charges $q_I$ are allowed to fluctuate at a fixed 
electric potential $\phi^I$ \cite{Ooguri:2004zv}:
\be
\label{zosv}
Z= \sum_{q_I\in\Lambda_e} 
\Omega(p^I,q_I)\ e^{\pi q_I \phi^I} := e^{{\cal F}(p^I,\phi^I)}
\ee
where $\Lambda_e$ is the lattice of electric charges in the large
volume polarization.
At leading order, using the $\CN=2$ attractor formalism one finds that 
the free energy ${\cal F}$ is expressed in terms of tree-level superpotential
$F$ via 
\be
\label{ftop}
{\cal F}(p^I,\phi^ I) = -\pi \Im F_0 ( X^I )
\ee 
where $X^I=p^I+i \phi^I$, so that the free-energy is identified as 
the modulus square of the topological wave function $\Psi=e^{i\pi F_0/2}$,
\be
e^{{\cal F}(p^I,\phi^I)} = \sum_{k^I \in \Lambda^*_e}
\Psi^*( p^I- i \phi^I - 2 k^I) \Psi( p^I+ i \phi^I + 2 k^I) 
\ee
The summation over $k^I$ on the right-hand side is necessary in order
to maintain the periodicity in imaginary integer shifts of $\phi^I$,
as follows from the quantization condition over $q_I$ in 
\eqref{zosv} \cite{Dabholkar:2005by,Dabholkar:2005dt}.
In particular, the Legendre transform of 
${\cal F}(p^I,\phi^ I)$ with respect to $\phi^I$ reproduces the leading
Bekenstein-Hawking entropy. In general, there are higher contributions from
worldsheet instantons and $R^2 F^{2h-2}$ higher-derivative interactions, 
but those are absent in $\CN=8$. There could be additional
contributions e.g. due to $R^4$ couplings, but their precise form is
not known at this stage. From the knowledge of the microscopic
degeneracies $ \Omega(p^I,q_I)$, one could in principle compute 
${\cal F}(p^I,\phi^ I)$ via \eqref{zosv}. Conversely, 
from the latter one can obtain the  microscopic 
degeneracies by Laplace transform,
\be
\label{osvii}
\Omega(p^I,q_I) = \int d\phi^I e^{{\cal F}(p^I,\phi^I) -\pi q_I \phi^I}
\ee
The integral \eqref{osvii} was evaluated in \cite{Dabholkar:2005by}
for classical prepotentials 
\be
\label{fi3}
F_0 = I_3( X^A )/ X^0
\ee
given by an arbitrary cubic polynomial
$I_3(X^A)$, in the absence of D6-brane charge ($p^0=0$).
In this section, we shall compute the integral \eqref{osvii}
for arbitrary charges, but for cases where $F_0$ is invariant under Legendre 
transform in all variables $X^0,X^A$. Remarkably, this property holds
in all cases of interest in this paper.

\subsection{Legendre invariant prepotentials and cubic integrals}

As shown in \cite{Gunaydin:1983bi},  homogeneous vector-multiplet
moduli spaces are classified by Jordan algebras $J$ of degree 3. 
In particular, their prepotential is of the form
\eqref{fi3}, where the homogeneous cubic polynomial 
$I_3(X^A)$ is the norm of $J$. As a consequence, $F_0$ is invariant
under a Legendre transform with respect to all variables at once\footnote{This
was not stated in this way in \cite{Gunaydin:1983bi}, but follows from
the axioms (M1-M5) in this paper.}. Independently, Legendre-invariant
homogeneous cubic polynomials in a finite number of variables have been
classified in \cite{etingof} (see also \cite{Pioline:2003uk}):
\begin{enumerate}
\item[(i)] $G=D_{n\geq 4}: I_3 = X^1 (X^2 X^3 + X^4 X^5+ \dots +
  X^{2n-6} X^{2n-5})$\ ;
\item[(ii)] $G=E_6: I_3 = \det( X )$, with $ X$ a $3\times 3$
  matrix;
\item[(iii)] $G=E_7: I_3 = \Pf ( X )$, with 
$ X$ an antisymmetric $6\times 6$ matrix;
\item[(iv)] $G=E_8: I_3 = X^3\vert_{\bf 1}$, with 
$X$ a {\bf 27} representation of $E_6$ and $I_3$ the singlet in 
the cubic power of  {\bf 27};
\item[(v)] $G=B_{n\geq 3}: I_3 = X^1 [ (X^2)^2+ X^3 X^4 + \dots +
  X^{2n-5} X^{2n-4}]$\ ;
\item[(vi)] $G=F_4: I_3 =  \det( X )$, with $ X$ a symmetric $3\times 3$
  matrix;
\item[(vii)] $G=G_2: I_3 =  X^3$, with $X$ a 
single variable.
\end{enumerate}
We have labeled each case by a group $G$, since the corresponding cubic
polynomial plays a crucial r\^ole in the minimal unitary representation
of $G$, as we shall review in Section 5.1.
Case (i) corresponds to the tree-level prepotential in $\CN=2$ heterotic
compactifications (the $n=4$ case corresponds to the $STU$ model), 
while we already encountered cases (ii-iv) in Section 1 of this paper. 

The assumption of invariance under
Legendre transform in all variables means that the solution to 
the equation $\nabla_X F_0(X)= Y$ is given by $X=\nabla_Y F_0(Y)$, i.e.
\be
\label{xy}
\left\{ \begin{matrix}
Y^A &=& -\partial_A I_3(X) / X^0\\
Y^0 &=& I_3(X) / (X^0)^2
\end{matrix} \right.
\quad
\Leftrightarrow
\left\{ \begin{matrix}
X^A &=& -\partial_A I_3(Y) / Y^0\\
X^0 &=& I_3(Y) / (Y^0)^2
\end{matrix} \right.
\ee
For $X$ and $Y$ related as in \eqref{xy}, we have
\be
\label{i3xi3y}
I_3(X) = [I_3(Y)]^2/(Y^0)^3\ ,\quad I_3(X)/(X^0)^3 = (Y^0)^3/I_3(Y)
\ee
hence
\be
\label{fsym}
F_0(X) + X^0 Y^0 + X^A Y^A = - F_0(Y)
\ee
This implies that the classical approximation to the 
Fourier transform of $\exp[iF_0(X)]$ is equal to $\exp[-iF_0(Y)]$.

\begin{table}
\begin{equation*}
\begin{array}{c@{\hspace{4mm}}c@{\hspace{4mm}}c@{\hspace{4mm}}c@{\hspace{4mm}}c@{\hspace{4mm}}c@{\hspace{4mm}}c@{\hspace{4mm}}c}
\hline
G & n_v & I_3 & H & H_0 & \kappa & \mu & \nu \\
\hline
D_{n\geq 4} & 2n-4 & X^1 C_{ab} X^a X^b & A_1 \times D_{n-2} & 
D_{n-4} & (-1)^n & n-1 & (n-4)/2\\
E_6 & 10 & \det(X) & A_5  & A_2 \times A_2 & -1 & 6 & 1/2\\
E_7 & 16 & \Pf(X)  & D_6  & A_5 & 1 & 9 & 1 \\
E_8 & 28 & 27^3\vert_1 & E_7 & E_6 & 1 & 15 & 2 \\
B_{n\geq 3} & 2n-3 & X^1 C_{ab} X^a X^b & A_1 \times B_{n-2} & 
B_{n-4} & (-1)^n & n-1 & (n-4)/2\\
F_4 & 7  & \det(X) &  C_3    &  A_2   & -8 & 3/2 & n.a.\\
G_2 & 2  & x^3     &  A_1    &  1   & 3  &  2 & n.a. \\
\hline
\end{array}
\end{equation*}
\caption{Data entering the construction of the minimal unipotent
representation of $G$, of functional dimension $n_v+1$. $I_3$ is
a homogeneous polynomial of degree 3 in $n_v-1$ variables, such that
$F=I_3(X)/X^0$ is invariant under Legendre transformation in all $n_v$
variables. The subgroup of $H\subset G$ acts by canonical transformations 
and $H_0\subset H$ by linear transformations of the variables. $\kappa$
is the numerical factor entering in the Hessian \eqref{dex}, and $\nu$
is the index of the Bessel function entering in the spherical vector
$f_K$ in \eqref{sph}. For $G=F_4,G_2$, the spherical vector does not
exist, as there is no $K$-singlet in the minimal representation.}
\end{table}

Let us now determine the 1-loop determinant. By explicit
computation, we find that the determinant of the Hessian 
of the map $X \to \nabla_X F$ in \eqref{xy} is equal to
\be
\label{dex}
\det[ \nabla_{X^I} \nabla_{X^J} F_0(X) ] = 
\left\{ 
\begin{array}{cc}
\kappa \ \left( \frac{I_3(X)}{(X^0)^3} \right)^{(n_v+2)/3}
\ ,\quad & G \neq B_n, D_n \\
\kappa \ \left( \frac{X^1}{X^0}\right)^{n_v-4} 
\ \left(\frac{I_3(X)}{(X^0)^3}\right)^{2} \ ,\quad & G = B_n, D_n 
\end{array}
\right.
\ee
where the number of complex variables  $n_v$ and the numerical factor
$\kappa$ can be read off in Table 1.
From general properties of the Legendre transform:
the Hessian at $Y=\nabla_X F$ is the inverse of the Hessian at
$X$: suppressing indices for simplicity,
\be
f(x) - x f'(x) = g (f'(x)) \Rightarrow -x f''(x) = f''(x) g'(f'(x))
\Rightarrow g'(f'(x))=x
\ee
Differentiating once more with respect to $x$ indeed gives
\be
g''(f'(x)) f''(x) = -1  \Rightarrow g''(y) = -1 /f''(x)
\ee
Thus, the determinant of the Hessian at the saddle point is
\be
\label{dey}
\det[ \nabla_{X^I} \nabla_{X^J} F_0(X) ]\vert_{Y} = 
\left\{ 
\begin{array}{cc}
\kappa \ \left( \frac{I_3(Y)}{(Y^0)^3} \right)^{-(n_v+2)/3}
\ ,\quad & G \neq B_n, D_n \\
\kappa \ \left( \frac{Y^1}{Y^0}\right)^{-(n_v-4)} 
\ \left(\frac{I_3(Y)}{(Y^0)^3}\right)^{-2} \ ,\quad & G = B_n, D_n 
\end{array}
\right.
\ee
This may also be obtained from \eqref{dex} using the identities
\eqref{i3xi3y}, and, in the $D_n$ case, $X^0/X^1=-Y^1/Y^0$.
This implies in the semi-classical (one-loop) approximation, for $G \neq D_n$,
\be
\begin{split}
\label{cubga}
\int dX^0 dX^A  (X^0)^\alpha [I_3(X)]^\beta
&\exp\left[i F_0(X) 
+ i (X^0 Y^0 + X^A Y^A) \right] \\
& \sim \kappa^{-1/2} \ 
(Y^0)^{\alpha'} [I_3(Y)]^{\beta'}
\exp\left[ -iF_0(Y) \right]
\end{split}
\ee
where 
\bse
\label{abp}
\bea
\alpha'&=& -2 \alpha -3 \beta - (n_v+2)/2 \\
\beta' &=& \alpha + 2 \beta + (n_v+2)/6 
\eea
\ese
In the $D_n$ case,
\be
\label{cubgad}
\begin{split}
\int & dX^0 dX^A  (X^0)^\alpha [I_3(X)]^\beta (X^1)^\gamma 
\exp\left[i F_0(X) 
+ i (X^0 Y^0 + X^A Y^A) \right] \\
& \sim \kappa^{-1/2} 
(Y^0)^{\alpha'} [I_3(Y)]^{\beta'} (Y^1)^{\gamma'} 
\exp\left[ -iF_0(Y) \right]
\end{split}
\ee
where
\bse
\label{abgp}
\bea
\alpha'&=& -2 \alpha -3 \beta - \gamma - (n_v+2)/2 \\
\beta' &=& \alpha + 2 \beta + \gamma + 1 \\
\gamma'&=& -\gamma + (n_v-4)/2
\eea
\ese
It is easy to check that the linear transformations \eqref{abp} 
and \eqref{abgp} are involutions, as it is necessary 
if the Fourier transform is to square to one.
In \cite{etingof}, it
was shown that, for special choices of $(\alpha,\beta,\gamma)$, 
the classical approximation is in fact exact\footnote{The idea
of the proof is to use the Mellin representation $e^{I_3/x^0}
= \frac{1}{2\pi i}
\int_{c-i\infty}^{c+i\infty} 
dz (-I_3/x^0)^{-z} \Gamma(z) dz$ and compute the integral over $x^I$
in terms of generalized Gamma functions.}: in particular,
for $\beta=0$ and $G\neq D_n$, 
\be
\label{cubg}
\begin{split}
\int dX^0 dX^A \ (X^0)^{-(n_v+2)/6} & \exp\left[i F_0(X) 
+ i (X^0 Y^0 + X^A Y^A) \right] \\
&\sim \kappa^{-1/2} 
(Y^0)^{-(n_v+2)/6}
\ \left(\frac{I_3(Y)}{(Y^0)^3}\right) 
\exp\left[ -iF_0(Y) \right]
\end{split}
\ee
or, in the $G=D_n$ case,
\be
\label{cubgd}
\begin{split}
\int dX^0 dX^A  \ (X^0)^{(2-n_v)/2} (X^1)^{(n_v-4)/2}
&\exp\left[i F_0(X) 
+ i (X^0 Y^0 + X^A Y^A) \right]\\
& = \kappa^{-1/2} \ 
(Y^0)^{-1} \exp\left[ -iF_0(Y) \right]
\end{split}
\ee
(This last identity can be checked by first doing the Gaussian
integral over $X^a$, then the integral over $X^0$ which yields a Dirac
distribution for the remaining $X^1$ integral).
These identities will prove very useful in evaluating the integral
\eqref{osvii}.

\subsection{Classical evaluation}
In this section, we evaluate the classical limit of the integral 
\eqref{osvii}, i.e. the Legendre transform of the free energy \eqref{ftop}
with respect to all electric potentials $\phi^I, I=0,\dots,n_v-1$.
For a prepotential $F_0$ given by \eqref{fi3}, independently of the
assumption of Legendre invariance, the free energy reads
\be
{\cal F} = \frac{\pi}{(p^0)^2+(\phi^0)^2}
\left\{
p^0 \left[ \phi^A \partial_A I_3(p) - I_3(\phi) \right]
+ \phi^0 \left[ p^A \partial_A I_3(\phi) - I_3(p) \right]
\right\}
\ee
In order to eliminate the quadratic term  in $\phi^A$ and reach a form
closer to \eqref{cubg}, it is convenient to change variables
\be
\label{xax0}
x^A= \phi^A - \frac{\phi^0}{p^0} p^A\ ,\quad
x^0= [(p^0)^2+(\phi^0)^2] / p^0
\ee
The entropy in the mixed ensemble becomes
\bea
{\cal S} &=& 
\langle {\cal F}(p^I,\phi^I) - \pi q_I \phi^I \rangle_{\{\phi^I\}}\\
&=& \pi \langle - \frac{I_3(x)}{x^0} 
+ \frac{\partial_A I_3(p) + p^0 q^A}{p^0} x^A
+ \frac{2 I_3(p) + p^0 p^I q_I }{p^0}\sqrt{\frac{x^0}{p^0}-1} 
\rangle_{\{x^I\}}
\eea
where the right-hand side should be extremized with respect to all
$\phi^I$ (recall that $I$ runs from 0 to $n_v-1$).
In order to get rid of the square root, it is convenient to introduce an
auxiliary variable $t$, and write
\be
{\cal S} = \pi \langle  - \frac{I_3(x)}{x^0} 
+ \frac{\partial_A I_3(p) + p^0 q^A}{p^0} x^A
-\frac{t}{4} 
\left(\frac{x^0}{p^0}-1\right) -  
\frac{(2 I_3(p) + p^0 p^I q_I)^2}{t\ (p^0)^2}  \rangle_{\{x^I,t\}} 
\ee
At fixed $t$, we recognize the Legendre transform of $F_0(x)=I_3(x)/x^0$
with respect to all variables $x^I$, at conjugate potentials
\be
\label{yay0}
y_A = \frac{\partial_A I_3(p) + p^0 q_A}{p^0}\ ,\quad
y_0 = -\frac{t}{4p^0}
\ee
Using the Legendre invariance of $F_0(x)$, we conclude that
the result of the extremization over $x^I$ is
\be
{\cal S} = \pi \langle
4 \frac{I_3(p^0 y)}{(p^0)^2 t}
-  \frac{[2 I_3(p) + p^0 p^I q_I]^2}{t\ (p^0)^2} 
- \frac{t}{4} \rangle_{t}
\ee
The extremization with respect to $t$ leads to $t=S_0/\pi$, 
where 
\be
\label{s0}
S_0 = \frac{\pi}{p^0} \sqrt{ 4 I_3(p^0 y) - [2 I_3(p) + p^0 p^I q_I]^2 }
\ee
finally leading to the classical entropy,
\be
\label{sclas}
{\cal S}(p^I,q_I) = S_0 
\ee
It is easy to check that \eqref{s0} is consistent with the general result
in \cite{Shmakova:1996nz} -- in fact, Legendre invariance is what
allows to solve Eq. (14) in \cite{Shmakova:1996nz} in closed form. 
Setting $F=I_3(X)/X^0$ as appropriate for the $G=E_8$ case,
and making use of \eqref{i4i3t}, we find that the classical entropy 
$S_0$ in \eqref{s0} is in fact the square root of the quartic 
invariant of $E_7$. Conversely, we find that the relation \eqref{i4i3t} 
between the quartic invariant $I_4$ and the cubic polynomial $I_3$
is a general consequence of the attractor formalism, and that
the entropy of $\CN=8$ black holes \eqref{kol} is controlled to leading
order by the prepotential $F=I_3(X)/X^0$.

The classical entropy \eqref{s0} can
be further simplified by making use once again of the Legendre invariance
of $F_0$. Applying \eqref{i3xi3y} to the case $X^A=\pa_A I_3(p), X^0=1$, 
we may expand
\be
\begin{split}
I_3( \partial_A I_3(p) + p^0 q_A) &=\\
&[I_3(p)]^2 + p^0 I_3(p) p^A q_A + (p^0)^2 \pa^A I_3(q) \pa_A I_3(p)
+ (p^0)^3 I_3(q)
\end{split}
\ee
which allows us to rewrite 
\be
S_0 = \pi \sqrt{4 p^0 I_3(q) - 4 q_0 I_3(p) + 4  \pa^A I_3(q) \pa_A I_3(p)
- ( p^0 q_0 + p^A q_A )^2}
\ee
reproducing \eqref{i4i3}.

We conclude that, to leading order, the topological amplitude controlling
the entropy of  $1/8$-BPS black holes is
\be
\label{psi}
\Psi(X) = \exp( I_3 (X^A)/ X^0 )
\ee
where $I_3$ is the cubic invariant of $E_6$. 15 of the complex
variables $X^A/X^0$ may be viewed as the K\"ahler classes of $H_2(\CX)$,
while the remaining 12 are a subset of the Narain moduli 
in \eqref{e7so66}.

\subsection{Beyond the classical limit}
The leading entropy \eqref{s0} above was the result of a tree-level 
saddle point approximation  to the integral \eqref{osvii}. It however receives
quantum corrections from fluctuations around the saddle point. In addition,
there may be corrections to the prepotential itself, although we have little 
control on them. In this section, we shall assume that $F_0$ is uncorrected,
and compute the micro-canonical degeneracies $\Omega^{(0)}(p,q)$ which
result from \eqref{osvii} under this assumption.

Performing the same change of variables as in \eqref{xax0}, 
and introducing the auxiliary variable $t$ by the usual Schwinger
representation, the OSV integral becomes 
\be
\begin{split}
\Omega^{(0)} &= \int \frac{1}{4\sqrt{\pi t}} \ dx^0 dx^A dt \\
&\exp\left[ \pi \left(
 - \frac{I_3(x)}{x^0} 
+ \frac{\partial_A I_3(p) + p^0 q^A}{p^0} x^A
-\frac{t}{4} 
\left(\frac{x^0}{p^0}-1\right) -  
\frac{(2 I_3(p) + p^0 p^I q_I)^2}{t\ (p^0)^2} \right)
\right] 
\end{split}
\ee
The integral over $x^0, x^A$ is now a Fourier transform
of the type \eqref{cubga}, with conjugate momenta $y_0,y_A$ given
in \eqref{yay0}. In the
saddle point approximation, we thus get, for $G\neq D_n$,
\be
\int \frac{\kappa^{-1/2}}{4\sqrt{\pi t}} dt \ 
\left(\frac{2^6 I_3(p^0 y)}{t^3}\right)^{\frac{n_v+2}{6}}
\exp\left[
4\pi \frac{I_3(p^0 y)}{(p^0)^2 t}
-  \pi\frac{[2 I_3(p) + p^0 p^I q_I]^2}{t\ (p^0)^2} 
- \frac{\pi t}{4}
\right]
\ee
while, for $G=D_n$,
\be
\int \frac{\kappa^{-1/2}}{4\sqrt{\pi t}} dt \ 
\left(\frac{2^6 I_3(p^0 y)}{t^3}\right) 
\left(\frac{4(\vec p^2 + p^0 q_1)}{t} \right)^{\frac{n_v-4}{2}}
\exp\left[
4\pi \frac{I_3(p^0 y)}{(p^0)^2 t}
-\pi  \frac{[2 I_3(p) + p^0 p^I q_I]^2}{t\ (p^0)^2} 
- \frac{\pi t}{4}
\right]
\ee
The remaining integral over $t$ is of Bessel type, with a saddle point
at $t= S_0/\pi$. The one-loop determinant 
$1/\sqrt{{\cal S}''(t^*)}=t^{1/2}$ cancels the
factor of $1/\sqrt{t}$ in front, leading to the result, for $G\neq D_n$,
\be
\label{omosv1}
\Omega^{(0)}(p^I, q_I) \sim [I_3(p^0 y)]^{(n_v+2)/6} S_0^{-(n_v+2)/2} e^{S_0}
\ee
or, in the $D_n$ case,
\be
\label{omosv2}
\Omega_{0}(p^I, q_I)  \sim I_3(p^0 y)\  (\vec p^2 + 2 p^0 q_1)^{(n_v-4)/2}
S_0^{-(n_v+2)/2} e^{S_0}
\ee
It is important to note that 
the pre-factors appearing in these expressions are 
inconsistent with U-duality \footnote{For $p^0\neq 0$ they 
mix the electric and magnetic charges, and the option of 
considering ratios at fixed electric charge,
as advocated in \cite{Dabholkar:2005by}, is no longer available.},
indicating that the naive flat integration measure in 
\eqref{osvii} is inconsistent with U-duality. In order to remedy this,
we may use the fact that, 
at the saddle point, the prefactors in \eqref{omosv1},\eqref{omosv2}
can be expressed in terms of the magnetic charges and electric potentials,
\bse
\bea
I_3(p^0 y) &=& -\frac14 |X^0|^6 
\left( C_{ABC} \Im t^A \Im t^B \Im t^C \right)^2\\
 \vec p^2  + 2 p^0 q_1 &=& |X^0|^2  \Im t^a C_{ab} \Im t^b
\eea
\ese
where
\be
X^I = p^I + i \phi^I \ ,\quad t^A = X^A / X^0
\ee
This should be compared to the standard expression for the K\"ahler potential
(see e.g. \cite{Moore:1998pn}, eq. 9.6)
\be
e^{-K} = -\frac43 |X^0|^2 C_{ABC} \Im t^A \Im t^B \Im t^C
\ee
In order to cure the non-U-duality invariance of \eqref{omosv1}, 
a possible option is thus 
to multiply the flat integration measure in \eqref{osvii}
by $e^{(n_v+3)K}/|X^0|^3$;
this will remove the first factor in \eqref{omosv1} while leaving
the power of $S_0$ untouched (a similar option holds for $G=D_n$). 
However, there is no guarantee that higher-loop corrections
would be U-duality invariant under this prescription.

A more attractive option is 
to use the additional relation, valid at the saddle point, 
\be
x^0=I_3(p^0 y)/(p^0 t^2)
\ee
The first factor in \eqref{omosv1} can therefore be removed by
multiplying the integration measure by $(p^0 x^0)^{(n_v+2)/6}$.
Denoting by $\Omega^{(1)}$ the result of this procedure, we have, 
in terms of the original variables,
\be
\label{osv1}
\Omega^{(1)} \sim \int d\phi^0 d\phi^A 
\left[ (p^0)^2 + (\phi^0)^2 \right]^{-(n_v+6)/2} 
e^{{\cal F} + \pi \phi^I q_I}
\ee
According to \eqref{cubg}, this has the great advantage of rendering 
the 1-loop approximation to the integral over $x^0,x^I$ exact. The remaining
$t$ integral becomes
\be
\Omega^{(1)} = \int \frac{\kappa^{-1/2}}{4\sqrt{\pi t}} 
dt\  t^{-\frac{n_v+2}{6}}
\exp\left[
4\pi \frac{I_3(p^0 y)}{(p^0)^2 t}
-\pi  \frac{[2 I_3(p) + p^0 p^I q_I]^2}{t\ (p^0)^2} 
- \frac{\pi t}{4}
\right]
\ee
leading to the manifestly U-duality invariant result, in the $G\neq D_n$ case,
\be
\Omega^{(1)} = \hat I_{(n_v-1)/6} (S_0) \sim  S_0^{(n_v+2)/6} e^{S_0}
\ee
Similarly, in the $D_n$ case, using the measure in 
\eqref{cubgd}, we get a universal result
\be
\Omega^{(1)} = \hat I_{1/2} (S_0) \sim S_0^{-1} \ e^{S_0} 
\ee
which agrees with the previous case for $G=D_4$. We should stress
that \eqref{osv1} is only an educated guess; it however meshes well
with the conjecture in the next section.

We may now compare this macroscopic prediction with the
microscopic counting: setting $G=E_8,n_v=28$ as appropriate
for case (iv), we find
\be
\Omega^{(1)} \sim \hat I_{9/2}(S_0)
\ee
On the other hand, if we believe that the attractor formalism should
only describe the 16 vector multiplets described by the prepotential
\eqref{f0n2}, the $G=E_7, n_v=16$ case (iii) applies, leading
to 
\be
\Omega^{(1)} \sim \hat I_{5/2}(S_0)
\ee
where $S_0$ is now proportional to the 
square root of quartic invariant \eqref{it4} 
of $SO(6,6)$. In either case,
the index of the Bessel function differs from the microscopic counting
in \eqref{smic}. This suggests that there may be logarithmic corrections to the
topological amplitude, or that the appropriate generating function
should have modular weight 7/2 (for $G=E_8$) or 3/2 (for $G=E_7$).
We leave this discrepancy as an open problem for further investigation.

\section{Black hole partition functions and theta series}
In the previous section, we have demonstrated that the $E_7$-invariant
entropy formula \eqref{kol}, including all 56 electric and magnetic
charges, follows from the attractor formalism based
on the prepotential \eqref{fi3} where $I_3(X)$ is the cubic invariant of 
$E_6$. On the other hand, we have mentioned that this prepotential lies
at the heart of the construction of the minimal representation 
of $E_8$ \cite{Kazhdan:2001nx,Gunaydin:2001bt}.
This suggests that $E_8(\IZ)$ may be a hidden symmetry of the partition 
function of 1/8-BPS black holes in 4 dimensions. In this section, we
try and flesh out this conjecture.

\subsection{Review of the minimal unipotent representation}
Let us start by a brief review of the construction of the minimal 
``unipotent'' representation of a simple Lie groups $G$ in the split 
(i.e. maximally non compact) real form  (see \cite{Kazhdan:2001nx} for 
more details, as well as \cite{Gunaydin:2000xr,Gunaydin:2001bt,Gunaydin:2004md,
Gunaydin:2005gd,Gunaydin:2005zz} for an equivalent approach using the formalism
of Jordan algebras). 

The minimal representation of a 
non-compact group $G$ is the 
unitary representation of smallest functional dimension \cite{Joseph:1974}
It can be obtained by quantizing the co-adjoint orbit of smallest 
dimension in $G$, i.e. the orbit of any root in the Lie algebra of $G$.
Without loss of generality, we consider the orbit of the
lowest root $E_{-\omega}$.
Under the Cartan generator $H_{\omega}=
[E_{\omega},E_{-\omega}]$, the Lie algebra of $G$ decomposes into
a graded sum of 5 subspaces,
\be
G=G_{-2} \oplus G_{-1} \oplus G_0 \oplus G_{+1} \oplus G_{+2}
\ee
where $G_{\pm 2}$ are one-dimensional vector spaces along the
highest/lowest root $E_{\pm\omega}$. $G_0$ further decomposes into 
a commuting sum $\IR \oplus H$, where the first summand
corresponds to the Cartan generator $H_{\omega}$.
Since $G_{-2}, G_{-1}$ and $H$ commute
with $E_{-\omega}$, the
co-adjoint orbit of $E_{-\omega}$ is generated by the action of the
$H_{\omega}\oplus G_{+1} \oplus E_{\omega}$. As any co-adjoint
orbit, it carries a $G$-invariant Kirillov-Konstant symplectic form, which
decomposes into a symplectic form on $G_{+1}$ and a symplectic
form on $H_{\omega} \oplus E_{\omega}$. This endows 
$G_{+1} \oplus E_{\omega}$ with the
structure of a Heisenberg algebra, whose central element is $E_{\omega}$. 
Furthermore, $G_0$ acts linearly on $G_{+1}$. Quantization proceeds
by choosing a Lagrangian submanifold $C$ in $G_{+1}$, and representing
the generators of $G$ as differential operators acting on the space of
functions on $\IR \times C $, where the first factor denotes 
the central element of the Heisenberg algebra 
$E_{\omega}$. A standard choice of Lagrangian
$C$ is to take the orbit under $G_1$ of $E_{\beta_0}$, where 
$\beta_0$ is the root attached to the affine root on the Dynkin
diagram of $G$ \cite{Kazhdan:2001nx}. 
Let $H_0$ be the subgroup of $H$ which commutes
with $E_{\beta_0}$.
Parameterizing $G_{+1}$ by coordinates $x_0,x_1,\dots
x_{n_v-1}$ and momenta $p^0,p^1,\dots, p^{n_v-1}$ (where $x_0$ is the
coordinate along $E_{\beta_0}$), one can show that this Lagrangian
manifold is defined by the set of equations
\be
\label{pi3x}
C = \left\{ (x_I, p^I)\in G_{+1} \ \vert \ 
p^I = \frac{\partial F_0}{\partial x_I}
\right\}\ ,\quad F_0= \frac{I_3 (x_A)}{x_0}
\ee
where $I_3(x_A)$ is the $H_0$-invariant cubic polynomial built
out of the $x_A$'s  (in mathematical terms, it is the relative invariant of the
regular prehomogeneous vector space associated to $H_0$, or the norm
of the Jordan algebra with reduced structure group $H_0$). 
In principle, the relation
\eqref{pi3x} may be rewritten a set of $H$-covariant homogeneous
constraints on the vector $(x_I, p^I)$. The invariance
of $I_3$ under Legendre transform implies that $C$ is invariant under
the exchange of all $x_I$ with $p^I$ at once: this is precisely the
action of a particular element $S$ in the Weyl group of $G$ (the longest
element in the Weyl group of $H_0$). Another Weyl element $A$ (the Weyl
reflection with respect to the root $\beta_0$) acts as a $\pi/2$
rotation in the $(y,x_0)$ plane. 

The result of this procedure is a unitary representation of $G$  
in the Hilbert space $\CH$ of functions of $n_v+1$ variables
$(y,x_I)$. Infinitesimal generators are represented 
by differential operators, of which we display a subset only:
\bea
E_{\omega} &=& y \ , \\
\quad E_{\beta_I} = y \partial_I&,&  E_{\gamma_I} = i x_I \\
H_{\beta_0} = -y \partial_y + x_0 \partial_0&,& 
H_{\omega} = -\mu - 2  y \partial_y - x^I \partial_I \\
E_{-\beta_0} = - x_0 \partial_y &+& \frac{i}{y^2} I_3(x^A) \\
E_{-\omega}=y p^2 +p(x_I p^I)+ x_0 I_3(p) &-& \frac{p^0}{y^2} I_3(x)
+ \frac{1}{y} 
\frac{\partial I_3(x)}{ \partial x_A}
\frac{\partial I_3(p)}{ \partial p^A}
\eea
where $p=i \pa/\pa y, p^A = i \pa/\pa x^A$ and $\mu$ is a numerical
constant displayed in Table 1. In the last
equation, we dropped the ordering terms for simplicity.
The Weyl reflections $S$ and $A$ are represented as
\bea
(S\cdot f) (y ,x_I) &= & \int dy_0 dy_A \ e^{i (x_I y^I)/y}
f(y ,y^I) \\
(A\cdot f) (y,x_0, x_A) &=& e^{-\frac{I_3(x_A)}{x_0 y}}
f \left( -x_0, y, x_A\right)
\eea
A vector of particular interest in the Hilbert space $\CH$ is the
spherical vector $f_K$, i.e. a function invariant under the maximal compact 
subgroup $K$ of $G$. The spherical vector has been 
computed for all simply laced groups $G$ in the split form
in  \cite{Kazhdan:2001nx}, and reads (for $G\neq D_n$)
\be
\label{sph}
f_K(\tilde X) = \frac{1}{|z|^{2\nu+1}} \hat K_{\nu} ( S_1 ) e^{-i S_2}
\ee
where $z=y+i x_0$, $\hat K_{\nu}(x)$ is related to the modified
Bessel function by $\hat K_{\nu}(x)= x^{-\nu} K_\nu(x)$, $\nu$
can be read off in Table 1, and 
\be
\label{s1s2}
S_1 = \sqrt{ 
\sum_{\alpha=0}^{n_v} \left[ \tilde X_\alpha^2 + 
(\nabla_{\alpha} \tilde F)^2 \right] }\ ,
\quad
S_2 = \frac{x_0\ I_3(x)}{y(y^2+x_0^2)}
\ee
and $\tilde X=( x_0, x_A, y)$ and 
\be
\tilde F_0(\tilde X) =\frac{I_3(x_A)}{\sqrt{y^2+x_0^2}}
\ee
For $G=D_n$, the spherical vector is instead
\be
\label{sphd}
f_K(\tilde X) = |z|^{-1}  \left(1+ \frac{x_1^2}{|z|^2}\right)^{(n-4)/2}
\hat K_{(n-4)/2} ( S_1 ) e^{-i S_2}
\ee
The term under the square root in \eqref{s1s2} is recognized as the 
squared norm 
of the vector $(\tilde X_\alpha,\nabla_{\alpha} \tilde F)$
in the Lagrangian submanifold $\IR \times C$, invariant under the
maximal compact 
subgroup of $H$. Expanding $S=S_1-iS_2$ in powers of $z$, we may
rewrite
\be
S= \frac{\sqrt{|z|^6 + |z|^4 \sum_A x_A^2 
+ |z|^2 \sum_A [\partial_A I_3(x_A)]^2 + [I_3(x_A)]^2}}{|z|^2}
- i \frac{x_0 \ I_3(x_A)}{y|z|^2}
\ee
In the limit $z\to 0$ with $I_3(x_A) >0$, 
the spherical vector \eqref{sph} therefore behaves as 
\be
\label{f0sub}
\log f_K \sim 
\frac{I_3(x_A)}{y z} - (\nu+\frac12) \log I_3(x_A) 
+ \frac12 I_3(x_A) \sum_A [\partial_A I_3(x_A)]^2
+ {\cal O}(|z|^2)
\ee
Using the spherical vector $f_K$ and a $G(\IZ)$-invariant distribution
$\delta_{G(\IZ)}$ in ${\cal H}^*$, 
we may now construct an automorphic theta series as
\be
\label{thde}
\theta_G( g) = \langle \delta_{G(\IZ)}, \rho(g) \cdot f_K \rangle
\ee
where $g$ takes value in $G(\IR)$ and $\rho(g)$ is the minimal unitary
representation of $G(\IR)$ in $\CH$ constructed 
above \cite{Kazhdan:2001nx,Pioline:2003bk}. Due to the 
invariance of $f_K$ under the maximal compact subgroup $K$ of $G$,
the left-hand side is a well-defined function on $G(\IR)/K$, which
is furthermore  invariant
under the arithmetic group $G(\IZ)$ -- in other words, an automorphic
form. Furthermore, the invariant distribution
$\delta_{G(\IZ)}$ can be obtained by adelic methods, and is equal to the
product over all primes $p$ of the 
spherical vectors over the p-adic fields
$\mathbb{Q}_p$ \cite{Pioline:2003bk}.

\subsection{The minimal representation of $E_8$}
Let us now spell out the above general construction for $E_8$
in more physical terms. $E_8$ is the U-duality group of type II
string theory compactified on $T^7$ (or M-theory compactified on $T^8$).
Since black holes are static solutions in 4 dimensions, it is natural
to consider black holes at finite temperature $T$, 
and think of the 4-th direction as a thermal circle of radius $R_0=1/T$.
In the decompactification limit to 4 dimensions, $E_8$ decomposes into
$E_7 \times Sl(2)$, where the second factor is generated by 
($E_{-\omega},H_{\omega},E_{\omega})$. Accordingly, the moduli space in 
3 dimensions factorizes into 
\be
\label{e83d}
\frac{E_8}{SO(16)} = \frac{Sl(2)}{U(1)} 
\times \frac{E_7}{SU(8)} \bowtie \IR^{56}
\ee
where the last factor transforms as a 56 representation under $E_7$.
Thus, there is a non-linear action of $E_8(\IR)$ on the 58-dimensional
space $Sl(2)/U(1) \times \IR^{56}$, by right multiplication on this 
decomposition (assuming that the fractions in \eqref{e83d} are left-cosets): 
this is the classical action of $E_8$ on the co-adjoint orbit 
of $E_{-\omega}$\footnote{By dropping the Cartan generator in $Sl(2)/U(1)$,
one obtains the ``quasi-conformal realization'' of $E_8$ on 
57 variables \cite{Gunaydin:2000xr}.}.
Using the general techniques\footnote{In a nutshell \cite{Obers:1998rn}: 
represent the 
root lattice in a basis where the fundamental roots are 
$e_{i+1}-e_i$ ($i=1,\dots 7$) and $e_1+e_2+e_3-e_0$ and associate
to any vector $\alpha=\sum_{I=0}^8 \alpha^I e_I$ 
the quantity $S=l_p^{3 \alpha_0}\prod_{i=1}^{8} R_i^{\alpha_i}$,
where $l_p$ is the 11-dimensional Planck length; 
if $\alpha$ a positive root, $S$ is the action of D-instanton
conjugate to a Peccei-Quinn modulus in $G/K$.} in \cite{Obers:1998fb}, 
is is easy to understand the
physical interpretation of these 58 variables: the first factor in
\eqref{e83d} is described by
\be
\label{yt}
y + i t = K_{0;01234567} + i R_0^2 V_{1234567}/l_p^9 
\ee
while $\IR^{56}$
is parameterized by two $Sl(8)$ antisymmetric 
matrices\footnote{We use the same notation as in \eqref{QPmat}, but
\eqref{QPmat} and \eqref{xpe8} are in fact conjugate to each other.}
$Q$ and $P$ (equation (4.71)
in \cite{Kazhdan:2001nx}, 
after flipping the last two rows and columns and relabelling
$R_8$ into $R_0$)
\be
\label{xpe8}
Q = \begin{pmatrix}
C_{0 ij} & C_{0i7} & K^i  \\
- C_{0i7} & 0 & K^7 \\
-K^i & - K^7 & 0
\end{pmatrix}\ ,\quad
P = \begin{pmatrix}
E_{0klmn7} & E_{jklmn7} & g_{0i}  \\
-E_{jklmn7}  & 0 & g_{07} \\
-g_{0i} & - g_{07} & 0
\end{pmatrix}
\ee
where $i,j$ run from 1 to 6, and a dualization over $(j)klmn$ is understood.
By decompactification of the thermal circle, the scalars $g_{I0},C_{IJ0}$,
$E_{IJKLM0}$,$K_{I;IJKLMNP0}$ ($I,J,\dots=1,\dots 7$)
become gauge fields in 4 dimensions,
which are precisely the 56 electric and magnetic gauge fields in \eqref{gf4}.
In fact, it is generally true that positive roots in the moduli space
are conjugate to instantons, which become black holes in one dimension 
higher \cite{Obers:1998fb,Ganor:1999uy}. 
This also allows to understand the meaning of $y,t$
in \eqref{yt}: the imaginary part is the product of the inverse temperature
square by the volume of the M-theory $T^7$ in Planck units. The
real part is the scalar dual of the Kaluza-Klein gauge field $g_{0\mu}$
in 3 dimensions. Thus, it is the potential conjugate to the
3-dimensional NUT charge, i.e. the first Chern class of the line bundle
of the time direction on the sphere at infinity\footnote{Angular momentum 
in 4 dimensions can be viewed as the dipole charge associated to the
NUT charge. In other words, spinning black holes may be obtained by
combining two stationary black holes with opposite non-zero NUT charges.}.

Now, in order to quantize this co-adjoint orbit, one should take a 
Lagrangian subspace in $\IR^{56}$. The standard polarization
described in Section 5.1 is obtained 
by Fourier transform over 
the last two columns (or rows) in $Q$, as well as $x^0$. 
Interpreting the direction $7$ as the M-theory direction the ``coordinates'' 
in this polarization consist of 
the 1+27 potentials $x_0=g_{07}, C_{ij0},  E_{ijklm0}, g_{0i}$ dual to the 
D0-brane, D2-brane, NS5-brane and Kaluza-Klein momentum on $T^6$. This
is precisely the ``large volume'' polarization in \eqref{lv}. In this
basis, the  cubic invariant of $E_6$ entering the prepotential \eqref{pi3x}
is given by
\be
I_3 = \Pf( [D2]^{ij} ) + \frac{1}{5!} \epsilon_{jklmnp} [kk]_i [D2]^{ij}
[NS]^{klmnp}
\ee
where again we identify charges with their conjugate potentials.
On the other hand, 
the $Sl(8)$-invariant polarization \eqref{xpe8} can be reached by 
Fourier transform over the 13 variables $[kk]_{i}$ and $[NS]^{klmnp}$.
The prepotential controlling the corresponding Lagrangian submanifold
is obtained by  Legendre transform of \eqref{fi3} over the same
variables, leading to 
\be
F_0^{Sl(8)} = \sqrt{ \Pf ( Q ) }
\ee
where $Q$ is the antisymmetric $8\times 8$ matrix in \eqref{xpe8}. This
is a useful hint on the spherical vector $f_{E_8}$ 
in the $Sl(8)$ polarization, which is unknown until now \cite{Kazhdan:2001nx}.

\subsection{Wigner function and spherical vector}
In order to properly formulate our conjecture, let us return to \eqref{osvii}:
as noticed in \cite{Ooguri:2004zv}, upon 
analytically continuing $\phi\to i\chi$, 
the left-hand side is interpreted as the Wigner function associated
to the topological wave function $\Psi=e^{F}$:
\be
\Omega(p^I,q_I) = \int d\chi \ \Psi^*(p^I - \chi^I)\ 
\Psi(p^I + \chi^I) \ e^{2\pi i \chi^I q_I}
\ee
Now, let us postulate that the microscopic degeneracies 
$\Omega(p^I,q_I)$ are invariant under $G(\IZ)$  ($E_7(\IZ) \subset G(\IZ)$
for M-theory on $T^7$), and investigate the 
consequences of this assumption for the wave function $\Psi$.
For illustration purposes, we shall consider $G=Sl(2,\IZ)$ 
acting on a single pair of conjugate charges $(p,q)$ as a doublet.
For the generator $q\to q+ a p$, writing
\be
\Omega(p,q+ap) = \int d\chi \ 
e^{-\frac{i\pi}{2} a (p-\chi)^2} \Psi^*(p- \chi)\ 
e^{\frac{i\pi}{2} a (p+\chi)^2} \Psi(p+ \chi) \ e^{i \chi q }
\ee
the right-hand side is identified as the Wigner function of
the transformed wave function
\be
\label{psita}
\tilde\Psi(p) = e^{\frac{i\pi a}{2} p^2}  \Psi(p) 
= \Psi(p) + \frac{i\pi a}{2}  p^2 \Psi(p) +
          {\cal O}(a^2)
\ee
Similarly, for an infinitesimal shift $q\to q+c p$, one may show
by integration by parts that
\be
\tilde\Psi(p) = \Psi(p) - \frac{i c}{8\pi} \  \partial_p^2  \Psi(p) 
+ {\cal O}(c^2)
\ee
Finally, under an exchange $(p,q)\to(-q,p)$, it is straightforward
to check that $\Psi(p)$ is mapped to its Fourier transform. This
means that, under a $Sl(2,\IR)$ linear transformation of the
phase space $(p,q)$, the wave function $\Psi(p)$ transforms  by a unitary
representation of $Sl(2,\IR)$ -- to wit, the metaplectic representation.
The $Sl(2,\IZ)$ invariance of the microscopic degeneracies 
$\Omega(p,q)$ is thus equivalent to the invariance of $\Psi$ under
$Sl(2,\IZ)$. 

In this simple case, this problem has a well known solution,
unique up to rescaling:
$\Psi(p)$ is simply the ``Dirac comb'' distribution $\delta_{\IZ}(p)=
\sum_{m\in \IZ} \delta(p-m)$. Indeed, since it is localized on the integers,
it is invariant\footnote{In fact, it is only invariant under \eqref{psita}
when $a\in 4\IZ$; this is due to the fact that the metaplectic group
is a 2-sheeted cover of $PSl(2,\IZ)$. This subtlety does not occur
when $G$ is simply-laced.}  under \eqref{psita}. It is also invariant under
Fourier transform by the Poisson resummation formula. 
Recall furthermore that it can be obtained as a product over all primes $p$
of the spherical vector of the metaplectic representation over $\mathbb{Q}_p$,
which is the function equal to 1 for $x\in \IZ_p$, 0 otherwise.
Setting
\be
\Psi(p) = \sum_{m\in \IZ} \delta(p-m)
\ee
we find that the Wigner function is
\be
\Omega(p,q) = \delta_{\IZ}(2p) \delta_{\IZ}(q)
\ee
which corresponds to a uniform distribution on the lattice of charges. 
Applying the prescription \eqref{thde} in this case
leads to the standard Jacobi theta series for $Sl(2,\IZ)$ 
\cite{Pioline:2003bk}.

\subsection{$\CN=8$ black holes in 4D and the $E_8$ theta series}

The lesson from the previous example is clear: assuming that the microscopic
degeneracies $\Omega(p^I,q_I)$ in M-theory compactified on $T^7$ are
indeed equal to the Wigner function of a wave function $\Psi$, the
latter has to be invariant under a unitary representation of $E_7(\IZ)$ 
acting on the space of 28 variables $p^I$. Unfortunately, the minimal
representation of $E_7$ has only functional dimension 17 (while the 
generic unitary representation of $E_7$, based on the coadjoint 
orbit of a generic diagonalizable element has functional dimension 61),
and it does not appear likely that $E_7$ have a unirep of dimension 28 (
although it does have a unirep of dimension 27 \cite{Gunaydin:2001bt}).
The minimal representation of $E_8$ however provides a natural unitary 
representation of $E_7$ on 28 variables, with an extra 
variable $y$, which is spectator under the action of $E_7$. 
Furthermore, the spherical vector
for this representation over $\mathbb{Q}_p$ is known for all primes,
providing a concrete $E_7(\IZ)$ (in fact $E_8(\IZ)$)
invariant distribution $\delta_{E_8(\IZ)}$. We thus propose that
the exact degeneracies (or rather, the helicity supertrace $\Omega_8$)
in M-theory compactified on $T^7$
are given by the Wigner transform of the distribution 
$\delta_{E_8(\IZ)}(y,p_A)$ in the $(y,p_A,q^A)$ space.
This proposal raises some interesting questions:
\begin{itemize}
\item[i)] The computation in Section 4.3 indicates that  the classical
limit of the Wigner function is effectively determined by the spherical vector
$f_K$ over $\IR$ rather than $\mathbb{Q}_p$.
It would be interesting to understand this in more detail.
\item[ii)] The spherical vector $f_K$ has subleading corrections \eqref{f0sub}
to the prepotential \eqref{fi3} as $z\to 0$. Can one interpret them 
as higher-derivative corrections to the prepotential \eqref{fi3} ?
\item[iii)] The spherical vector is annihilated by the compact generators
$E_{\alpha}\pm E_{-\alpha}$. Can we understand these partial
differential equations, especially when $\alpha$ is the highest
root, as a $\CN=8$ version of the holomorphic anomaly equations ?
\item[iv)] One could in principle compute the Wigner function
in the full phase space $(t,y,p^A,q_A)$, which would be invariant
under the full $E_8(\IZ)$ symmetry. Can this distribution be understood
as a black hole partition function at finite temperature and 
NUT potential ?
\end{itemize}
We hope to return to these questions in a future publication.

\subsection{$\CN=4$ black holes in 4D and the $D_{16}$ theta series}
A similar reasoning can be applied in $\CN=4$ models such as type IIA
string theory compactified on $K3\times T^2$, or its dual the heterotic
string compactified on $T^6$. A counting function was proposed long
ago in \cite{Dijkgraaf:1996it}, based on an automorphic form
of the modular group $Sp(4,\IZ)$ of genus-2 Riemann
surfaces\footnote{In
the proposal \cite{Dijkgraaf:1996it}, the $SO(6,22,\IZ)$ symmetry is realized
trivially by including dependence on the square of the inner
products of the charges $q_e^2,q_m^2,q_e\cdot q_m$ only. This may
not be true when the charges have some common divisors.}, and
recently rederived using the 4D/5D connection in \cite{Shih:2005uc}.
Compactifying down to 3D dimensions, the U-duality group 
$Sl(2,\IZ) \times SO(6,22)$ is enhanced to $SO(8,24,\IZ)$, while
the moduli space decomposes as
\be
\label{6223d}
\frac{SO(8,24)}{SO(8)\times SO(24)} = 
\frac{Sl(2)}{U(1)} \times \left[ \frac{Sl(2)}{U(1)}
\times\frac{SO(6,22)}{SO(6)\times SO(22)} \right] \times
\IR^{56}
\ee
Again, the last factor in \eqref{6223d} can be identified as the time
component of the 28+28 electric and magnetic gauge fields in 4 dimensions,
conjugate to the 28+28 electric and magnetic charges.
It transforms linearly
as a (2,28) representation of the 4-dimensional U-duality
group. The first factor corresponds to the same field as in \eqref{yt}.
By right multiplication, the 4-dimensional group acts symplectically
on $Sl(2)/U(1)\times \IR^{56}$, the coadjoint orbit of the lowest root
of $SO(8,24)$. The minimal representation of $SO(8,24)$ is obtained by
quantizing this orbit, and acts on functions of 29 variables: for
the standard $SO(6,22)$-invariant polarization, based on the prepotential
\be
F_0 = X_1 X^a C_{ab} X^b /X^0
\ee
where $C_{ab}$ is a signature (5,21) quadratic form, these are
the 28 electric charges in 4 dimensions, together with 
the variable $y$ conjugate to the 3D NUT charge.
It should be straightforward to adapt
the $SO(16,16)$ spherical vector \eqref{sphd} 
to the $SO(8,24)$ real form\footnote{The minimal representation
of $SO(4,28)$ real form
of $D_{16}$ has been constructed recently in \cite{Gunaydin:2005zz}.}. 
The $SO(8,24,\IZ)$
invariant distribution $\delta_{D_{16}(\IZ)}$ may be computed 
as before by tensoring the spherical vectors over all $p$-adic fields
computed in \cite{sasha}. We thus propose that
the micro-canonical degeneracies in the heterotic string 
compactified on  $T^6$
are given by the Wigner function of the distribution $\delta_{D_{16}(\IZ)}$.
It would be very interesting to understand the relation with the formula
proposed in \cite{Dijkgraaf:1996it}.

\subsection{Conformal quantum mechanics}

Finally, we would like to mention an interesting interpretation of the
minimal representation, as the spectrum-generating symmetry of a 
a conformal quantum mechanical system \cite{Gunaydin:2000xr,Pioline:2002qz}. 
Consider the
universal $Sl(2)$ subgroup generated by $(E_{\omega},H_{\omega},
E_{-\omega})$ in the standard polarization. Performing a 
canonical transformation \cite{Pioline:2002qz}
\bea
y=\frac12 \rho^2\ &,& x_A = \frac12 \rho q_A \\
p=\frac{1}{\rho} \pi - \frac{1}{\rho^2}q_A \pi^A\ &,&
p^A = 2\frac{\pi^A}{\rho}
\eea
the highest root generator $E_{\omega}$ becomes (up to computable
ordering terms) the Hamiltonian of a De Alvaro Fubini Furlan-type
quantum mechanical system \cite{deAlfaro:1976je}:
\be
\label{cosmoe}
E_{-\omega} = \frac12 \pi^2 + \frac{I_4(\pi^A, q_A)}{2\rho^2}
\ee
where $I_4(\pi^A, q_A)$ is given by the same expression \eqref{i4i3}
which related the black hole entropy to the cubic prepotential. 
The universal $Sl(2)$ factor is interpreted as the conformal group
in 0+1 dimensions, and is only part of the full $E_8$ spectrum 
generating symmetry. The spherical vector $f_K$ may be viewed 
is the ``most symmetric'' state, which is as close to the
ground state as one may hope to get for a Hamiltonian whose
spectrum is unbounded both from below and from above. It would be very
interesting to understand the relation between this 
quantum mechanical system and the one controlling the 
cosmological / attractor flow 
of the moduli in the near-horizon geometry introduced 
in \cite{Ooguri:2005vr}. The conformal quantum mechanics \eqref{cosmoe}
 may also be related to the conformal models introduced in 
\cite{Claus:1998ts,Gibbons:1998fa,Michelson:1999dx}.

\section{Conclusion}
In this paper, we discussed the degeneracies of 4D and 5D BPS black holes
in maximally supersymmetric compactifications of M-theory or type II string
theory, with U-duality as a powerful tool. Using the 4D/5D lift, we
computed the exact degeneracies of 4D black holes with D0,D2 and unit
D6 charge, and found agreement with the general expectation from U-duality
at leading order. We also proposed a natural generalization of the
4D/5D lift to include all 56 charges of $\CN=8$ supergravity in 4 dimensions.
Utilizing the remarkable invariance of the
prepotential under Legendre transform, we computed to leading order
the ``topological amplitude'' which controls the $\CN=8$ attractor
formalism, and found an hint of a $E_8$ hidden symmetry in the black
hole partition function. By analysing the physical interpretation of
the minimal unipotent representation of $E_8$, we conjectured that
exact BPS black hole degeneracies
should be given by the Wigner function
of the unique $E_8(\IZ)$-invariant distribution in this representation.
A similar conjecture relates the degeneracies of $\CN=4$ black holes
to the minimal representation of $SO(8,24)$. The spherical vectors
are known explicitely in both cases, and it would be very interesting
to test these conjectures against other approaches such 
as \cite{Dijkgraaf:1996it,Shih:2005uc}. 

Another interesting question is the relation
of the $E_8$ conformal quantum mechanics \eqref{cosmoe}
which underlies the minimal representation
with the radial/cosmological flow investigated in \cite{Ooguri:2005vr}:
in particular, one would like to know if the $E_8$ conformal
quantum mechanics \eqref{cosmoe} admits a supersymmetric extension,
and if so, whether the truncation to BPS states is equivalent to
the invariance under the maximal compact subgroup. If so, this would 
indicate that the ``wave function of the Universe'' in this mini-superspace
formulation is indeed the spherical vector $f_K$,
as suggested in \cite{Pioline:2002qz}.

Assuming that the admittedly speculative conjectures in this paper hold
true, it is interesting to ask about the generalization to $\CN=2$
supersymmetry. Several years ago, M. Kontsevitch made the ``very wild guess'' 
that the topological string amplitude should be an infinite dimensional
solution to the ``master equation''  
${\rm Fourier}(e^F)=e^{{\rm Legendre}(F)}$ \cite{maxim}, of which
the cubic prepotentials $F=I_3(X)/X^0$ which we
encountered in this work are finite-dimensional solutions.
Since the topological amplitude $\Psi=e^F$
can be thought of as a wave function in the topological B-model
\cite{Witten:1993ed}, it is indeed
natural to expect that symplectic transformations
on the Calabi-Yau periods will act by Fourier transform, and relate
Gromov-Witten instanton series in different geometric phases. It is our
hope that a careful study of the $\CN=8$ case will help in making
these ideas more precise.

\acknowledgments
It is a pleasure to thank A. Dabholkar, F. Denef, 
R. Dijkgraaf, G. Moore, N. Obers, A. Strominger, E. Verlinde
and A. Waldron for valuable discussions or correspondence, and
especially M. Kontsevitch for providing me with the notes of his
1995 Arbeitstagung lecture. 

\vspace*{1cm}
\noindent {\it Historical notes:} 
(i) While this manuscript was being written up,
a preprint appeared which independently derived the main result 
of Section 3 \cite{Shih:2005qf}. I am grateful to A. Strominger for sending 
me a draft prior to publication.
(ii) I also wish to thank M. Gunaydin for many helpful remarks on an 
earlier version of this manuscript, and for pointing out the relation
to very special supergravities and Jordan algebras. 
(iii) In the original manuscript, the extra charge was misidentified
as the angular momentum in 4 dimensions. I realized and corrected this mistake 
in late August 2005, after the article was published. 
In the present version, the extra charge is correctly identified 
as the NUT charge in 3 dimensions. A forthcoming paper will elaborate 
at length on this and other issues \cite{gnpw}.
(iv) Following the recent paper \cite{Pestun:2005}, the 6 moduli 
which promote the 9 complex moduli of $U(3,3)/U(3)\times U(3)$ to
$SO^*(12)/U(6)$ in \eqref{f0n2} are now understood as generalized Calabi-Yau
moduli. 

\appendix

\section{$E_7$ minimal representation and black holes in 5 dimensions}
By the same reasoning as above, one may also expect that the black
hole partition function in 5 dimensions may be related to the 
minimal representation of $E_7$, since this is the U-duality group
which appears under compactification on a thermal circle
to 4 dimensions. The minimal representation of $E_7$ is based on the
decomposition
\be
\label{e73d}
\frac{E_7}{SU(8)} = 
\frac{Sl(2)}{U(1)} \times \frac{SO(6,6)}{SO(6)\times SO(6)} 
\bowtie \IR^{32}
\ee
This is different from the decomposition
$E_7\to E_6\times \IR$ which controls the decompactification limit
to 5 dimensions, and which is instead related to the ``conformal'' realization
of $E_7$ on 27 variables \cite{Gunaydin:2000xr}. 
Nevertheless, as we shall see, it may be sufficient to 
describe the Ramond-Ramond charges in 5 dimensions. Using the
same techniques as before, we identify the last factor in
\eqref{e73d} as the 16+16 Ramond-Ramond gauge fields and scalars
in Type IIA on $T^5$ (where $R_1$ is the M-theory circle, $R_{2,3,4,5,6}$
are the radii of $T^5$ and $R_7$ is the radius of the 6th direction): 
in the $SO(5,5)$ polarization (Eq. (4.55) in \cite{Kazhdan:2001nx}),
the 5+10+1 ``coordinates'' correspond to the 5D scalars
\be
\label{spx}
Q = \left\{ g_{1i}, C_{ijk}, C_{123456} \right \}
\ee
(where $i,j,k$ run from 2 to 6) 
while the 1+10+5 ``momenta'' correspond to the reduction of the
5D RR vectors along the 6th direction,
\be
P = \left\{ g_{17}, C_{ij7}, C_{ijklm7} \right\}
\ee
In addition, the $Sl(2)/U(1)$ factor corresponds to $y+i t = 
C_{234567} + i V_{234567} /l_p^6$.
This is not the ``standard'' polarization of \cite{Kazhdan:2001nx}, which is
invariant under $Sl(6)$, the mapping class group of type 
IIB string theory compactified on the T-dual 
$T^6$. The latter can however be reached by a Fourier transform
over the 5 variables $g_{17}, C_{237}, C_{247}, C_{257},
C_{267}$ \cite{Kazhdan:2001nx}.

While $E_7$ is expected to unify 5D black holes and 
5D black strings \cite{Bena:2004tk}, we find that the minimal representation 
of $E_7$ is unsuitable for this purpose, as it unifies 5D black holes and 5D 
instantons. Nevertheless, it may turn out to be relevant for 5D black
degeneracies, in the following sense: $E_7$ admits a maximally commuting
algebra of dimension 27, transforming as a 27 representation of $E_6$,
which contains the $SO(5,5)$ spinor $Q$ in \eqref{spx} as an isotropic vector
(a ``pure 27-sor'' in the terminology of \cite{Pioline:2003bk}, i.e.
a solution of the quadratic equations $27 \otimes 27\vert_{\bar{27}}=0$).
It is natural to conjecture that the Fourier coefficients of the $E_7$ 
theta series with respect to this 
commuting algebra may have a relation to the degeneracies of
small black holes with zero tree-level entropy, i.e. 
solutions to the cubic equation $27^3|_1=0$.

\section{$E_6$ minimal representation and black holes in 6 dimensions}
Similarly, we expect that black hole degeneracies 
in 6 dimensions may have a hidden
$E_6(\IZ)$ symmetry, larger than the naive  U-duality group $SO(5,5,\IZ)$.

The minimal representation of $E_6$ follows from the decomposition
\be
\label{e6dec}
\frac{E_6}{USp(8)} = \frac{Sl(2)}{U(1)} \times \frac{Sl(6)}{SO(6)} 
\bowtie \IR^{20}
\ee
and acts on functions of 11 variables,
which can be identified as
\be
Q = \begin{pmatrix}
0 & C_{345} & C_{245} & C_{235} & C_{234} \\
  & 0       & C_{145} & C_{135} & C_{134} \\
  &         & 0       & C_{125} & C_{124} \\
  & a/s     &         & 0       & C_{123} \\
  &         &         &         & 0
\end{pmatrix}
\ ,\quad
y = E_{123456} 
\ee
together with their conjugates
\be
P = \begin{pmatrix}
0 & C_{126} & C_{136} & C_{146} & C_{156} \\ 
  & 0       & C_{236} & C_{246} & C_{256} \\ 
  &         &    0    & C_{346} & C_{356} \\ 
  & a/s     &         & 0       & C_{456} \\
  &         &         &         & 0
\end{pmatrix}
\ ,\quad
t = V_{123456}/l_p^6
\ee
in the $Sl(5)$-invariant polarization (\cite{Kazhdan:2001nx}, Eq. (4.45),
after permuting a permutation (13)(45) on the rows and columns). This
is related to the ``standard'' $Sl(3)\times Sl(3)$ invariant polarization,
by prepotential $F_0 = \det(X)/X^0$, by Fourier transform over
$C_{126},C_{136},C_{236}$. Choosing $R_6=1/T$ as the radius of the 
thermal circle, the variables $P$ and $y$ can be interpreted as the
electric potentials dual to the $[M2]^{IJ}$ ($I,J=2,\dots 6$)
and $[M5]$ black hole charges, leaving no room for the $[KK]_I$ charges.

As in the $E_7$ case, the decomposition \eqref{e6dec} does not preserve
the U-duality symmetry $SO(5,5)$ in 6 dimensions\footnote{Instead,
the branching $\IR\times SO(5,5)\subset E_6$ leads to the conformal
realization of $E_6$ on 16 variables, where the 16 variables transform
as a spinor of $SO(5,5)$ \cite{Gunaydin:2000xr}.}. Nevertheless, it can
be checked that the 11 charges $[M2]^{IJ}$ and $[M5]$ transform as
an isotropic vector of $SO(5,5)$ -- in other words, a pure spinor
of $SO(5,5)$, which satisfies $16 \otimes 16\vert_{\overline{ 16}}=0$. 
It us thus tempting to conjecture that the Fourier coefficients
of the $E_6$ theta series with respect to this dimension 16 Abelian
subalgebra are related to degeneracies of ``small'' black holes in 6
dimensions (indeed, all BPS black holes in 6 dimensions are ``small'',
in that a smooth solution of the Einstein-Maxwell equations with 
the required charges does not exist \cite{Klebanov:1996un,Cvetic:1996dt}). 
It would be very interesting to understand the relation with
the approach in \cite{Dijkgraaf:1996cv,Dijkgraaf:1996hk}.


\providecommand{\href}[2]{#2}\begingroup\raggedright\endgroup

\end{document}